\documentclass[fleqn,10pt]{wlscirep}
\usepackage[utf8]{inputenc}
\usepackage[T1]{fontenc}
\usepackage{mathtools}
\usepackage{float}
\usepackage{amsmath}
\usepackage{comment}
\usepackage{bookmark}
\title{Discontinuous phase transitions in the multi-state noisy $q$-voter model: quenched vs. annealed disorder.}

\author[1]{Bart\l omiej Nowak}
\author[1]{Bartosz Sto\'n}
\author[1,*]{Katarzyna Sznajd-Weron}
\affil[1]{Department of Theoretical Physics, Faculty of Fundamental Problems of Technology, Wrocław University of Science and Technology, 50-370 Wrocław, Poland}

\affil[*]{katarzyna.weron@pwr.edu.pl}

\begin{abstract}
    We introduce a generalized version of the noisy $q$-voter model, one of the most popular opinion dynamics models, in which voters can be in one of $s \ge 2$ states. As in the original binary $q$-voter model, which corresponds to $s=2$, at each update randomly selected voter can conform to its $q$ randomly chosen neighbors only if they are all in the same state. Additionally, a voter can act independently, taking a randomly chosen state, which introduces disorder to the system. We consider two types of disorder: (1) annealed, which means that each voter can act independently with probability $p$ and with complementary probability $1-p$ conform to others, and (2) quenched, which means that there is a fraction $p$ of all voters, which are permanently independent and the rest of them are conformists. We analyze the model on the complete graph analytically and via Monte Carlo simulations. We show that for the number of states $s>2$ the model displays discontinuous phase transitions for any $q>1$, on contrary to the model with binary opinions, in which discontinuous phase transitions are observed only for $q>5$. Moreover, unlike the case of $s=2$, for $s>2$ discontinuous phase transitions survive under the quenched disorder, although they are less sharp than under the annealed one.  
\end{abstract}
\begin{document}

\flushbottom
\maketitle
%
%
\thispagestyle{empty}

\section{Introduction}
It might seem that determining the type of a given phase transition is interesting only from the physics point of view. However, it has been reported that the hysteresis appears in real social systems \cite{Sche:Wes:Bro:03,Bis:15,Str:Liz:17,Pru:etal:18,Cen:etal:18}, which means that this issue is important also from the social point of view. Because hysteresis cannot appear within the continuous phase transition, researchers working in the field of opinion dynamics try to determine conditions under which discontinuous phase transitions appear \cite{Nyc:Szn:Cis:12,Vie:Cro:16,Che:etal:17,Tuz:Fer:Equ:18,Enc:etal:18,Enc:etal:19,Now:Szn:19,Abr:Paw:Szn:19,Chm:etal:20}.

In this paper we focus on two factors that are known to influence the type of transition, namely the type of disorder (quenched vs. annealed) and the number of states. It is known, that discontinuous phase transitions can be rounded (become less sharp) or even totally forbidden in the presence of the quenched disorder \cite{Aiz:Weh:89,Bor:Mar:Mun:13,Vil:Bon:Mun:14,Jed:Szn:17}. On the other hand, 
the larger number of states supports discontinuous phase transitions. The classical example is the Potts model: in two dimensions discontinuous phase transitions are observed for the number of states larger than $4$, whereas continuous ones for the smaller values of states \cite{Wu:82}. Similar situation has been reported for the majority-vote model. For the binary model only continuous phase transitions are observed \cite{Enc:etal:19, Vie:Cro:16, Vil:Mor:09}, whereas for more than two states the model undergoes a discontinuous order-disorder phase transition \cite{Li:etal:16,Oes:Pir:Cro:19}.

In this paper we introduce the generalized version of the noisy $q$-voter model, in which each agent can be in one of several discrete states, similarly as it was done already for the linear voter \cite{Red:19,Kha:Gall:20,Vaz:Los:Bag:19,Boh:Gro:12,Her:Gal:19,Sta:Bar:Sat:12}, majority-vote \cite{Li:etal:16,Oes:Pir:Cro:19,Che:Red:05,Mel:etal:10,Vil:etal:20,Che:Li:18} or other models of opinion dynamics \cite{Ban:Mal:19,SznWer:Szn:Wer:20}. We show that already for the $3$-state model only discontinuous phase transitions are possible. Moreover, we consider the model under two approaches, the quenched and the annealed one, and we show that discontinuous phase transitions can survive under the quenched disorder, similarly as in \cite{Net:Bri:20}. Only for the binary opinions, which were studied originally, the quenched disorder forbids discontinuous phase transition \cite{Jed:Szn:17}. 

\section{The model}
In this paper we propose a generalization of the original binary $q$-voter model (qVM) with independence \cite{Nyc:Szn:Cis:12}, known also as the noisy nonlinear voter \cite{Per:etal:18} or the noisy $q$-voter model \cite{Vie:etal:20}. Therefore, we consider a system of $N$ agents placed in the vertices of an arbitrary graph. In this paper we will focus on the complete graph, for which exact analytical calculations can be provided. In the generalized model, each agent $i$ is described by a dynamical $s$-state variable $\sigma_{i}(t) \in \{0,1,2,3,\dots,s-1\}$. As in the original $q$-voter model \cite{Cas:Mun:Pas:09}, which corresponds to $s=2$, a voter can be influenced by its neighbors only if the group of $q$ agents, chosen randomly out of the neighborhood of a given voter, is unanimous. Additionally, a voter can change its opinion to a random one, independently of others, as proposed by Nyczka et al. \cite{Nyc:Szn:Cis:12}. 

These two competitive processes -- conformity to others (ordering) and  independence (disordering), were originally introduced as alternatives appearing with complementary probabilities $1-p$ and $p$, respectively. Such an annealed approach led to two types of phase transitions in the original  $q$-voter model: continuous for $q \le 5$ and discontinuous for $q>5$. Later on, it was shown that replacing the annealed disorder by the quenched one reduced all transitions to continuous ones \cite{Jed:Szn:17}.

\begin{figure}[ht!]
\centering
\includegraphics[width=0.9\linewidth]{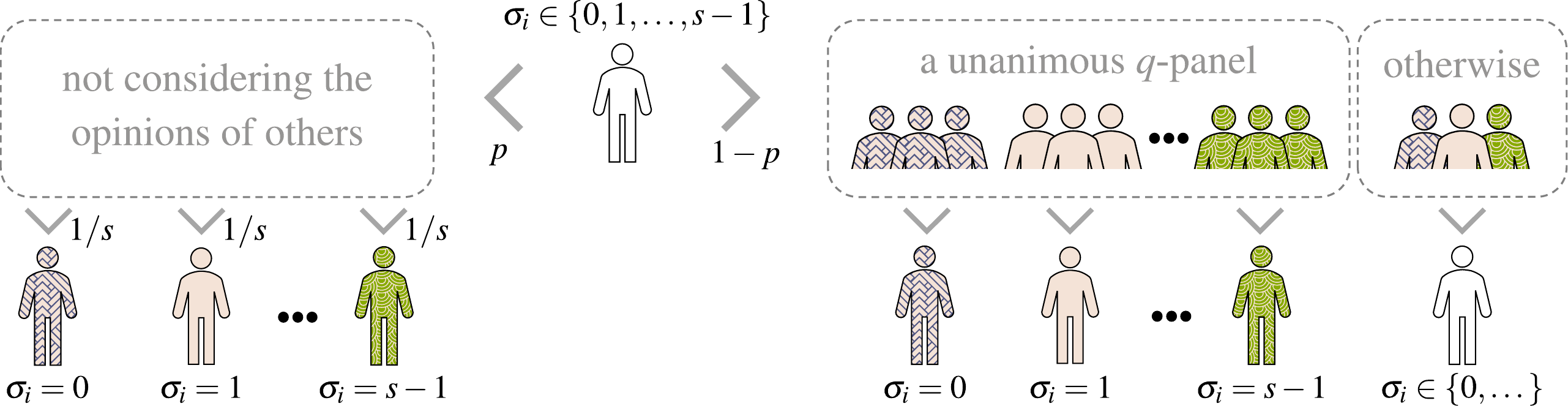}
\caption{Visualization of an elementary update for the multi-state $q$-voter model with independence. Within the annealed approach two alternative social responses, independence and conformity, appear with complementary probabilities $p$ and $1-p$. Whereas, within the quenched approach, a fraction $p$ of agents is permanently independent, whereas others are always conformists.}
\label{fig:model}
\end{figure}

In this paper we consider both types of disorder, annealed and quenched, and corresponding elementary updates are the following:
\begin{itemize}
    \item \textbf{Annealed approach:} \begin{enumerate}
                    \item site $i$ is randomly chosen from the entire graph,
                    \item a voter at site $i$ acts independently with probability $p$, i.e. changes its opinion to randomly chosen state (each state can be chosen with the same probability $1/s$),
                    \item with complementary probability $1-p$ a group of $q$ neighbors is randomly selected (without repetitions) and if all $q$ neighbors are in the same state, the voter at site $i$ copies their state.
                    \end{enumerate}
    \item \textbf{Quenched approach:} \begin{enumerate}
                \item site $i$ is randomly chosen from the entire graph,
                \item if the voter is independent (a fraction $p$ of all agents is permanently independent), then it changes its opinion to randomly chosen state (each state can be chosen with the same probability $1/s$),
                \item if the agent is conformist (a fraction $1-p$ of all agents is permanently conformists), a group of $q$ neighbors is randomly selected (without repetitions) and if all $q$ neighbors are in the same state, the voter at site $i$ copies their state.
                \end{enumerate}
\end{itemize}
As usually time is measured in Monte Carlo Steps (MCS), and a single time step consists of $N$ elementary updates, visualized in Fig. \ref{fig:model}. It means that one time unit corresponds to the mean update time of a single individual.

\section{Methods}
In this section, we are going to analyze the annealed and quenched formulations of the multi-state $q$-voter model (MqVM). We use both the analytical as well as the Monte Carlo approach. We focus on the mean-field description of the model, which corresponds to the fully connected graph. This approach was already applied to various binary-state \cite{Nyc:Szn:Cis:12,Jed:Szn:17,Jed:Szn:20,Mor:etal:13} and multi-state \cite{Vaz:Los:Bag:19,Li:etal:16} dynamics. We are aware that Monte Carlo (MC) simulations can be carried out only for the finite system, whereas analytical results correspond to the infinite one. However, it occurs that already for systems of size $N = 10^{5}$ simulation results overlap the analytical ones. 

The main goal of our study is to check how the number of states and the type of disorder influence the phase transition, observed in the original $q$-voter model with independence \cite{Nyc:Szn:Cis:12}. Therefore, we need to find the relation between stationary values of the concentration $c_{\alpha}$ of agents with a given opinion $\alpha=0,1,2,3,\dots,s-1$ and model's parameters $p$ and $q$. The concentration $c_{\alpha}$ is defined as:
\begin{equation}
    c_{\alpha} = \frac{N_{\alpha}}{N},
\end{equation}
where $N_{\alpha}$ denotes the number of agents with opinion $\alpha$. As usually, concentrations of all states sum up to one:
\begin{equation}
    \sum_{\alpha = 0}^{s-1}c_{\alpha} = \frac{N_0 + N_1 + \dots + N_{s-1}}{N} = 1.
    \label{eq:sumtoone}
\end{equation}
 
Based on the values of $c_{\alpha}$ we distinguish the following phases:
\begin{itemize}
    \item The disordered phase, in which all opinions are equinumerous i.e. $c_{0} = c_{1} = \dots = c_{s-1} = \frac{1}{s}$. 
    \item The ordered phase, in which one or more opinions dominate over the others. A special case within this phase is the state of consensus, i.e. when all voters share the same opinion $c_{\alpha} = 1, c_{\beta} = c_{\gamma} = \dots = 0$. 
    \item The coexistence phase (possible only in case of discontinuous phase transitions), if both ordered and disordered phases can be reached depending on the initial state of the system. 
\end{itemize}

Our model is based on the random sequential updating, i.e. in a single update only one agent can change its state. Thus, the concentration $c_{\alpha}$ can increase or decrease by $1/N$ or remain constant with the respective probabilities:
\begin{align}
\gamma^{+} (s,c_{\alpha},q,p) &= Prob \left\{c_{\alpha} \rightarrow c_{\alpha} + \frac{1}{N}\right\} \equiv \gamma_{\alpha}^{+}, \nonumber \\ 
\gamma^{-} (s,c_{\alpha},q,p) &= Prob \left\{c_{\alpha} \rightarrow c_{\alpha} - \frac{1}{N}\right\} \equiv \gamma_{\alpha}^{-}, \nonumber \\ 
\gamma^{0} (s,c_{\alpha},q,p) &= Prob \left\{c_{\alpha} \rightarrow c_{\alpha}\right\} = 1-\gamma^{+} (s,c_{\alpha},q)-\gamma^{-} (s,c_{\alpha},q) \equiv 1-\gamma_{\alpha}^{+} - \gamma_{\alpha}^{-}.
\end{align}
The dynamics of our model in the mean-field limit is given by the rate equation:
\begin{equation}
    \frac{dc_{\alpha}}{dt} = \gamma_{\alpha}^{+} - \gamma_{\alpha}^{-} = F(s,c_{\alpha},q,p), 
    \label{eq:dcdt}
\end{equation}
where $F(s,c_{\alpha},q,p)$ can be interpreted as the effective force acting on the system \cite{Nyc:Szn:Cis:12,Cas:Mun:Pas:09}. 
\subsection{Annealed approach}
Within the annealed approach a system is homogeneous, i.e. all agents are identical and transition rates can be expressed as:
\begin{align}
    &\gamma_{\alpha}^{+} =  \sum_{i \ne \alpha} P(i)\left[(1-p)P^{q}(\alpha|i) + \frac{p}{s}\right], \nonumber \\ 
    &\gamma_{\alpha}^{-} =  \sum_{i \ne \alpha} P(\alpha)\left[(1-p)P^{q}(i|\alpha) + \frac{p}{s}\right],
    \label{eq:gammas}
 \end{align}
where $P(i)$ is the probability of choosing a voter in $i$-th state and $P(\alpha|i)$ is the conditional probability of picking a neighbor in state $\alpha$ given that a target voter is in state $i$. Inserting $\gamma_{\alpha}^{\pm}$ to Eq. (\ref{eq:dcdt}) we obtain:
\begin{equation}
    F(s,c_{\alpha},q,p) =  \sum_{i \ne \alpha}\left[P(i)\left((1-p)P^{q}(\alpha|i) + \frac{p}{s}\right) -  P(\alpha)\left((1-p)P^{q}(i|\alpha) + \frac{p}{s}\right)\right].
\end{equation}
When events of picking a voter in state $i$ and a neighbor in state $\alpha$ are independent, which is true in case of a complete graph, then $P(\alpha|i)=P(\alpha)$. If we additionally assume that $\forall_{\alpha} P(\alpha) = c_{\alpha}$, which is also true for a complete graph, we end up with the simple formula:
\begin{equation}
    F(s,c_{\alpha},q,p) = \frac{dc_{\alpha}}{dt} =  \sum_{i \ne \alpha} \left[c_{i}\left((1-p)c_{\alpha}^{q} + \frac{p}{s}\right) -  c_{\alpha}\left((1-p)c_{i}^{q} + \frac{p}{s}\right)\right].
    \label{eq:dcsdt}
\end{equation}
Stationary states of Eq. (\ref{eq:dcdt}) are those for which 
\begin{equation}
    \frac{dc_{\alpha}}{dt} = F(s,c_{\alpha},q,p) = 0.
    \label{eq:dcsdt0}
\end{equation}
\begin{figure}[ht!]
    \centering
    \includegraphics[width=\linewidth]{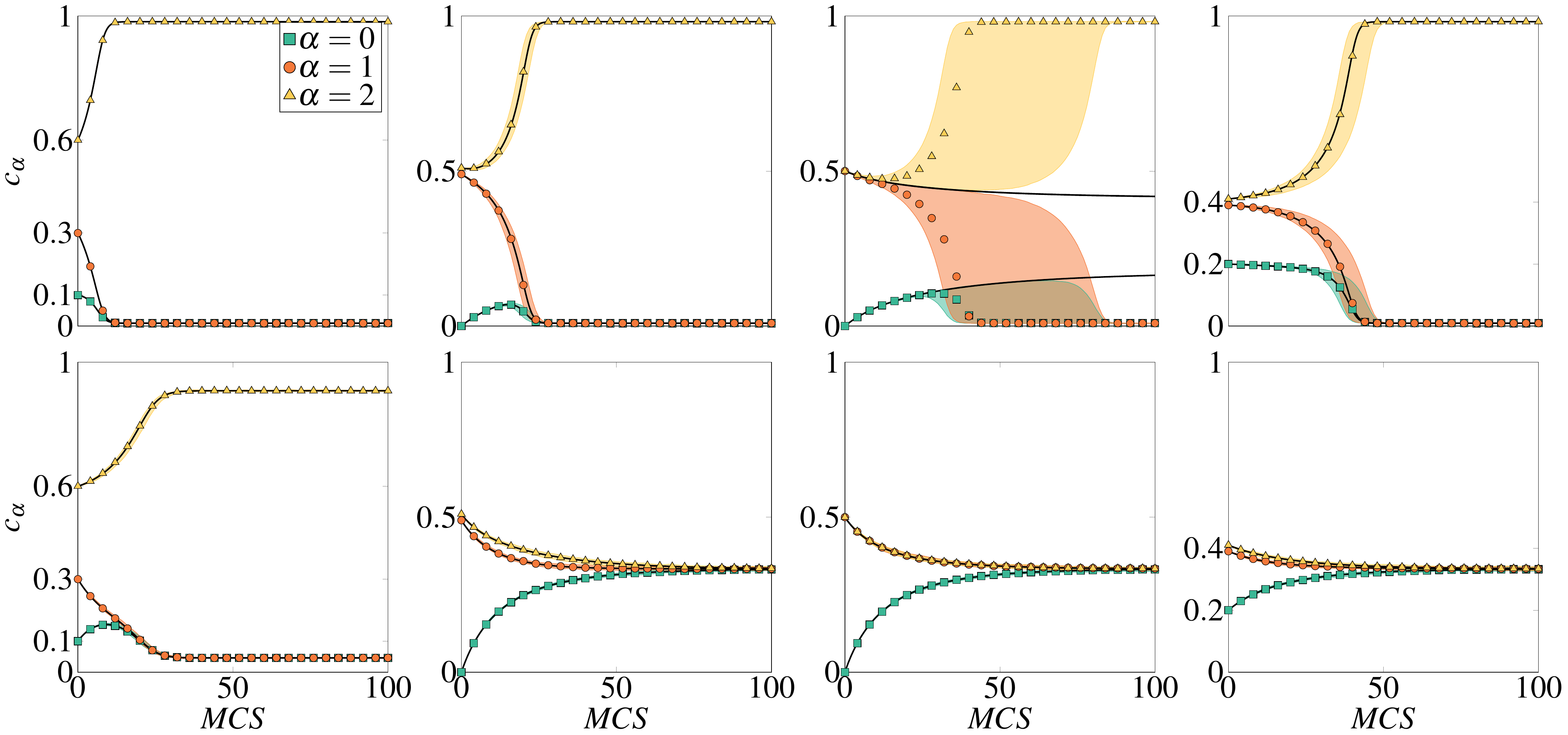}
    \caption{Trajectories for the multi-state annealed $q$-voter model on the complete graph of size $N=10^5$ with the size of the influence group $q=5$ and $s=3$ states. Upper and lower panels differ by the amount of noise: $p=0.025$ at the top row and $p=0.09$ at the bottom one. Markers and color areas represent the outcome of Monte Carlo simulations and thick solid black lines are results of analytical prediction obtained from Eq. (\ref{eq:dcsdt}). Symbols represent median trajectory over $50$ samples. The shaded color areas show the range of trajectories, i.e. are limited by 0 and 100th quantiles. Note that Monte Carlo simulations show a good agreement with analytical solutions in all panels, except of the third one in the upper row. The reason for this inconsistency is that in this case we are dealing with the hyperbolic (saddle) fixed point, i.e., a stable, as well as an unstable manifold exist \cite{Str:94}. Therefore,  within MC simulations the system always eventually leaves such a state due to the finite-size fluctuations.}
    \label{fig:trajec}
\end{figure}
The obvious solution of the above equation, which is valid for arbitrary value of $p$, is $c_{0} = c_{1} = \dots = c_{s-1} = \frac{1}{s}$. The other solutions can be obtained by solving numerically equation (\ref{eq:dcsdt0}). However, independently we can provide also the general analytical solution based on the analogy to the Potts model \cite{Wu:82}. In our model opinions are equivalent and there is no external field so we deal with $Z_s$ symmetry that can be broken due to the noise. Because we have such a noise, introduced by independence, we expect an order-disorder phase transition.  At the critical value of the noise (temperature in the Potts model and independence here), the $Z_s$ symmetry is broken and the system choose spontaneously one of $s$ states as a dominant one. From the mathematical point of view, such a transition corresponds to the  bifurcation, at which fixed point $c_{0} = c_{1} = \dots = c_{s-1} = \frac{1}{s}$ looses stability \cite{Str:94}. However, because it is a fixed point, although unstable, if initially the system is exactly at this point, it will stay in this point forever. It shows how important the initial state is, if we analyze a system of an infinite size.

If initially one opinion dominates over the others, i.e. $c_{\alpha}(0) > c_{\beta}(0), c_{\gamma}(0), c_{\delta}(0), \dots$, where $\alpha, \beta, \gamma, \delta, \dots \in \{0,1,\dots,s-1\}$ the system reaches an absorbing state in which this opinion still dominates over the others and the concentrations of all the others are equal $c_{\alpha} > c_{\beta} = c_{\gamma} = c_{\delta} = \dots $, see Fig. \ref{fig:trajec}. 
Similarly, if initially two or more equinumerous states dominate over the others the system reaches an absorbing state in which concentrations for these states are still equal and larger than the concentrations of others:
\begin{align}
    c_{\alpha} = c_{\beta} &> c_{\gamma} = c_{\delta} = c_{\epsilon} = \dots \nonumber \\
    c_{\alpha} = c_{\beta} &= c_{\gamma} > c_{\delta} = c_{\epsilon} = \dots \nonumber \\
    c_{\alpha} = c_{\beta} &= c_{\gamma} = c_{\delta} > c_{\epsilon} = \dots \nonumber \\
    &\vdots \nonumber \\
    c_{\alpha} = c_{\beta} &= c_{\gamma} = c_{\delta} = c_{\epsilon} = \dots = \frac{1}{s}.
\label{eq:gen_sol}
\end{align}
It means that in the final state at most two values of opinion's concentrations are possible. This observation, together with the normalizing condition (\ref{eq:sumtoone}) indicates that all solutions can be written in terms of a single variable $c$, which describes the concentration of a one given state. Because in our model all states are equivalent, we can choose any of them as a representative one. Therefore, let us denote the concentration of state $0$ by $c$ and then the concentrations of all remaining states can be expressed with $c$ by using condition (\ref{eq:sumtoone}):
\begin{align}
    &c_{0} = \dots = c_{s-(\xi + 1)} = c, \nonumber \\  
    &c_{s-\xi} = \dots = c_{s-1} = \frac{1-(s-\xi)c}{\xi},
    \label{eq:cxi}
\end{align}
where $\xi = 1,2,\dots,s-1$ and $\xi=0$ indicates solution, where all states are equinumerous: $c_{0} = c_{1} = \dots = c_{s-1} = \frac{1}{s}$. 

\begin{figure}[ht!]
    \centering
    \includegraphics[width=\linewidth]{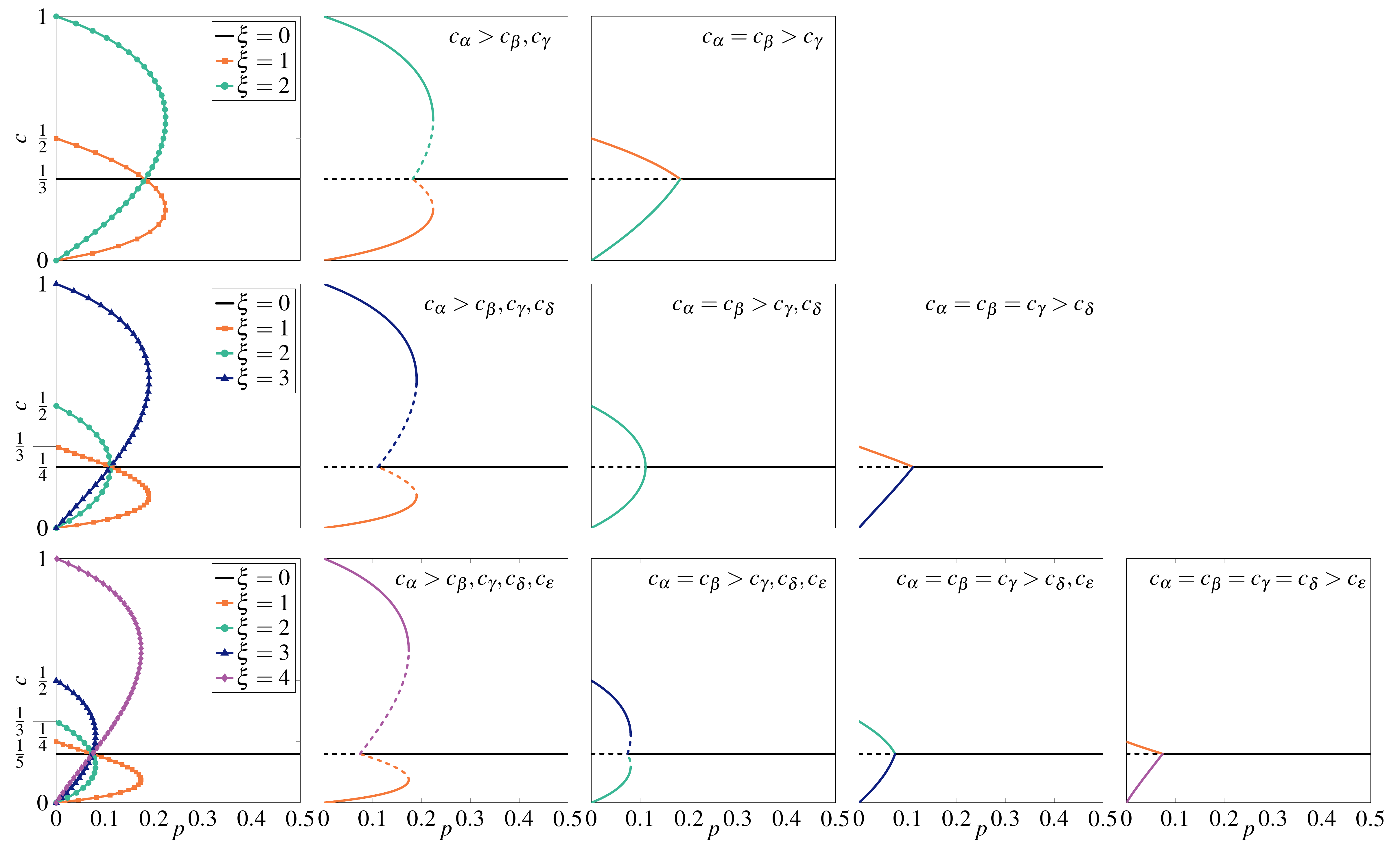}
    \caption{Steady states, given by the solution of Eq. (\ref{eq:pxiannealed}), for the annealed model with $q = 3$. Each row corresponds to a different number of states: $s=3$ (top panels), $s=4$ (middle panels), $s=5$ (bottom panels). The first column represents all possible solutions indexed with different values of $\xi$ without the distinction between the stable and unstable solutions. The remaining columns represent stationary states for  initial conditions indicated in the top right corner of each panel, where $\alpha, \beta, \gamma, \delta, \epsilon \in \{0,1,\dots,s-1\}$. Stable solutions are denoted by the solid lines, whereas the unstable ones are marked with the dashed lines.}
    \label{fig:xieqannealed}
\end{figure}

 Inserting Eq. (\ref{eq:cxi}) to Eq. (\ref{eq:dcsdt}) we obtain
\begin{equation}
    F(s,c,q,\xi) = (1-p)\left[(1-(s-\xi)c)c^q - c\xi \left(\frac{1-(s-\xi)c}{\xi}\right)^q\right] + \frac{p}{s}(1-sc).
\end{equation}
The stationary solutions different than $c = \frac{1}{s}$ are not that easy to derive in the simple form $c = c(p)$. However, since above equation is linear with the parameter $p$, we can derive the opposite relation from $F(s,c_{st},q,\xi) = 0$, i.e. $p = p(c_{st})$ \cite{Nyc:Szn:Cis:12}:
\begin{equation} 
    p = \frac{s\left[c_{st}^{q}-(s-\xi)c_{st}^{q+1}-\xi c_{st} \left(\frac{1-(s-\xi)c_{st}}{\xi}\right)^{q}\right]}{s\left[c_{st}^{q}-(s-\xi)c_{st}^{q+1}-\xi c_{st} \left(\frac{1-(s-\xi)c_{st}}{\xi}\right)^{q}\right] - 1 + c_{st}s}.
    \label{eq:pxiannealed}
\end{equation}
For $s=2$ and $\xi = 1$ the above equation correctly  reproduces the analytical solution for the original binary $q$-voter model with noise \cite{Nyc:Szn:Cis:12}. 
For more states, namely $s>2$, the above relation produces $s-1$ stationary solutions for $\xi = 1,2,\dots,s-1$ respectively, see Fig. \ref{fig:xieqannealed}. 

The information about the stability of these states is given by the sign of the first derivative of the effective force with respect to the concentration $c$ at the steady point:
\begin{equation}
    F'(s,c,q,\xi) = \left.\frac{dF(s,c,q,\xi)}{dc}\right|_{c = c_{st}}.
\end{equation}
The state is stable if $F'(s,c_{st},q,\xi) < 0$ and unstable if  $F'(s,c_{st},q,\xi) > 0$. Based on this analysis, two critical points can be identified: $p=p_{1}^{*}$ in which solution $c_{st} = 1/s$ loses stability (so called a lower spinodal) and $p = p_{2}^{*}$ in which steady state given by Eq. (\ref{eq:pxiannealed}) loses stability (so called an upper spinodal). 

At $c_{st} = 1/s$ we can determine the stability analytically, i.e. we are able to derive a formula for the lower spinodal. To do so we calculate the derivative of the effective force
\begin{equation}
    F'(s,c,q,\xi) = (1-p)\left[
        c^{q-1}q(1-c(s-\xi)) - c^{q}(s-\xi) - \xi \left(\frac{1-c(s-\xi)}{\xi}\right)^{q} + (s-\xi)cq\left(\frac{1-c(s-\xi)}{\xi}\right)^{q-1}
    \right] - p,
\end{equation}
which for $c_{st} = \frac{1}{s}$ gives
\begin{equation}
    F'\left(s,\frac{1}{s},q,\xi\right) = (1-p)\left(\frac{1}{s}\right)^q s (q-1) - p.
\end{equation}
From the above equation we see that $c_{st} = \frac{1}{s}$ is stable for $p>p_{1}^{*}$ and unstable otherwise, where
\begin{equation}
    p_{1}^{*} = \frac{q-1}{q-1+s^{q-1}}.
    \label{eq:annealedspinodal}
\end{equation}
The same result can be obtained in several different ways \cite{Nyc:Szn:Cis:12,  Now:Szn:20},  for example by taking the limit $c\rightarrow 1/s$ in Eq. (\ref{eq:pxiannealed}). As expected, for  $s=2$ the result for $p_{1}^{*}$ agrees with the one for the original $q$-voter model with independence \cite{Nyc:Szn:Cis:12}. We see in Eq. (\ref{eq:annealedspinodal}) that, the transition point depends on the size of the influence group $q$ and number of states $s$. However, it does not depend on the value of $\xi$, which means  that all stationary solutions intersect in the same point $p_{1}^{*}$, as clearly seen in Fig. \ref{fig:xieqannealed}.

\begin{figure}[ht!]
	\centering
	\includegraphics[width=\linewidth]{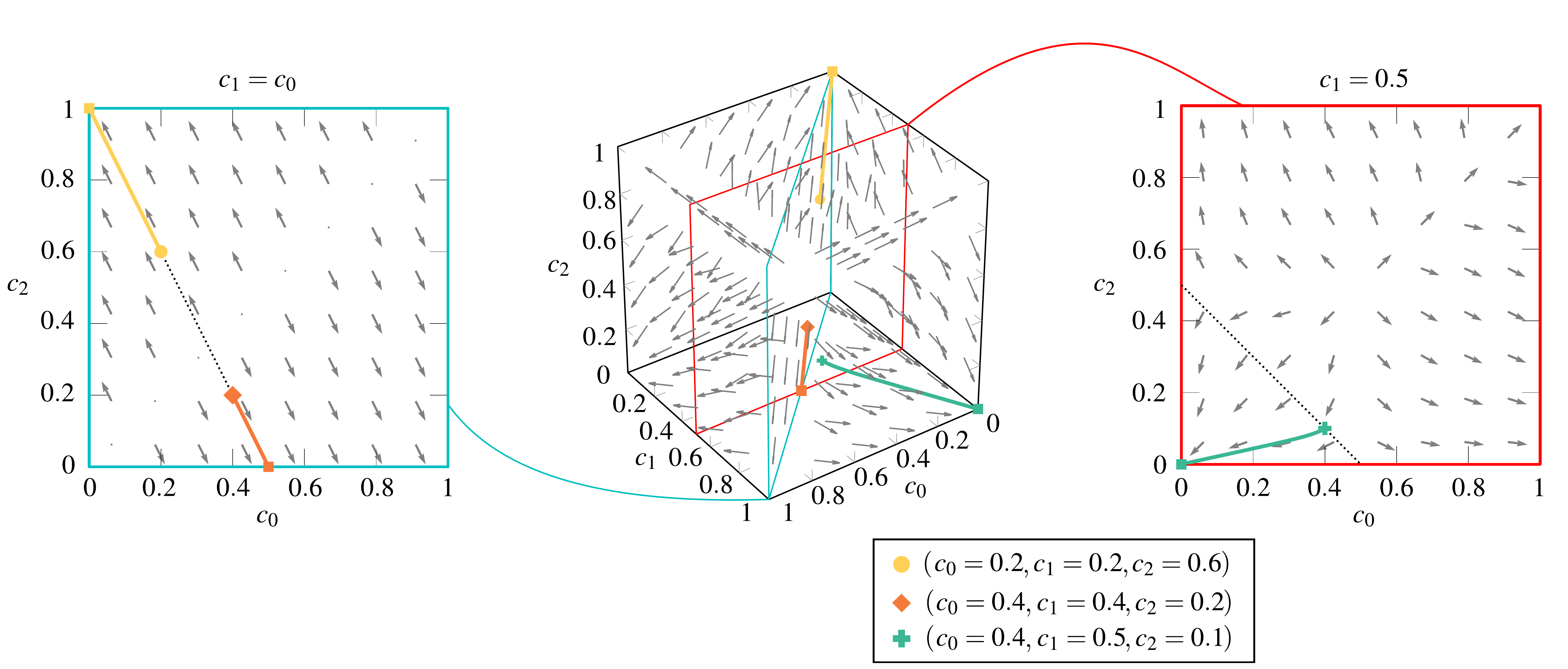}
	\caption{Flow diagrams for the annealed model with $s=3$ states, the group of influence $q=5$ and probability of independence $p=0$  obtained from Eq. (\ref{eq:dcsdt}). Arrows indicate the direction of the flow in the system. Squares refer to stationary points, whereas other markers represent initial points, as indicated in the legend. Lines that connect them represent trajectories. Space for clarity is presented for $c_{0}, c_{1}, c_{2} \in [0,1]$ as independent variables, whereas dotted lines on insets represent possible initial conditions which fulfill condition $c_{0} + c_{1} + c_{2} = 1$.}
	\label{fig:flow}
\end{figure}

The stability of other solutions of Eq. (\ref{eq:pxiannealed}) can be determined numerically. In Fig. \ref{fig:flow} we present the flow diagram for $s = 3$ and the noise parameter $p = 0$ as an example. It is visible that the states with only one dominant opinion are attractive. It means that from almost all initial conditions the system reaches the stationary state in which one opinion significantly dominates over the others. However, also another type of solution, namely the hyperbolic (saddle) \cite{Str:94} fixed point appears with more than one dominating opinion. In this case a stable, as well as an unstable manifold exist: the point is reached only from the initial state in which two or more equinumerous opinions dominate over the others but it cannot be reached from any other state. This type of solution has been observed also for the  multi-state majority-vote model \cite{Li:etal:16}. 

The steady state related to the saddle point in which several equinumerous opinions dominate over the others is visible only within the analytical approach but not within the MC simulations. In the latter case, the system initially seems to go towards the saddle point. However, after some time fluctuations push the system into the attractive steady state with only one dominant opinion, as shown in the third (from left) upper panel of Fig. \ref{fig:trajec}.

\subsection{Quenched approach}
Under the quenched approach, we have two types of agents \cite{Jed:Szn:17}: independent and conformists. For each type we introduce the concentration of agents in a given state,  similarly as for the annealed model. The only difference in respect to the annealed approach is that this time we consider separately $c_{(\mathbf{I},\alpha)}$ and $c_{(\mathbf{C},\alpha)}$ for independent and conformist voters in state $\alpha$, respectively. As a result the total concentration of voters in state $\alpha$ is
\begin{equation}
    c_{\alpha} = pc_{(\mathbf{I},\alpha)} + (1-p) c_{(\mathbf{C},\alpha)}.
    \label{eq:quenchedcalpha}
\end{equation}
Therefore, now the mean-field dynamics is given by two equations instead of one:
\begin{align}
    &\frac{dc_{(\mathbf{I},\alpha)}}{dt} = F_{\mathbf{I}}(s,c_{(\mathbf{I},\alpha)},q,p), \\
    &\frac{dc_{(\mathbf{C},\alpha)}}{dt} = F_{\mathbf{C}}(s,c_{(\mathbf{C},\alpha)},q,p).
\end{align}
Similarly as for the annealed approach we have
\begin{align}
    &F_{I}(s,c_{(\mathbf{I},\alpha)},q,p) = \gamma_{\mathbf{I}}^{+} (s,c_{(\mathbf{I},\alpha)},q) - \gamma_{\mathbf{I}}^{-} (s,c_{(\mathbf{I},\alpha)},q) \equiv \gamma_{(\mathbf{I},\alpha)}^{+} - \gamma_{(\mathbf{I},\alpha)}^{-}, \\
    &F_{\mathbf{C}}(s,c_{(\mathbf{C},\alpha)},q,p) = \gamma_{\mathbf{C}}^{+} (s,c_{(\mathbf{C},\alpha)},q) - \gamma_{\mathbf{C}}^{-} (s,c_{(\mathbf{C},\alpha)},q) \equiv \gamma_{(\mathbf{C},\alpha)}^{+} - \gamma_{(\mathbf{C},\alpha)}^{-},   
\end{align}
 where $\gamma_{(\mathbf{I},\alpha)}^{+}$ and $\gamma_{(\mathbf{I},\alpha)}^{-}$ are probabilities that the number of independent agents in state $\alpha$ increases and decreases respectively in a single update. The probabilities $\gamma_{(\mathbf{C},\alpha)}^{+}$ and $\gamma_{(\mathbf{C},\alpha)}^{-}$ describe the same, but for conformist agents. These probabilities can be express analogously as in the annealed approach:
\begin{align}
   &\gamma_{(\mathbf{I},\alpha)}^{+} =  \sum_{i \ne \alpha}\frac{P_{\mathbf{I}}(i)}{s},\\
   &\gamma_{(\mathbf{I},\alpha)}^{-} =  \sum_{i \ne \alpha}\frac{P_{\mathbf{I}}(\alpha)}{s},\\ 
   &\gamma_{(\mathbf{C},\alpha)}^{+} =  \sum_{i \ne \alpha}P_{\mathbf{C}}(i)P^{q}(\alpha|i),\\
   &\gamma_{(\mathbf{C},\alpha)}^{-} =  \sum_{i \ne \alpha}P_{\mathbf{C}}(\alpha)P^{q}(i|\alpha),
\end{align}

where $P(i)$ is the probability of choosing a voter with $i$-th state, $P_{\mathbf{I}}(i)$/$P_{\mathbf{C}}(i)$ is the probability of choosing a independent/conformist voter with $i$-th state and $P(\alpha|i)$ is the conditional probability of picking the neighbor in state $\alpha$  given that a target voter is in state $i$. 

As previously, $P(\alpha|i)=P(\alpha)$, and $\forall_{\alpha} P(\alpha) = c_{\alpha}$, and $\forall_{\alpha} P_{\mathbf{I}}(\alpha) = c_{(\mathbf{I},\alpha)}$, $\forall_{\alpha} P_{\mathbf{C}}(\alpha) = c_{(\mathbf{C},\alpha)}$, for the complete graph. Therefore:
\begin{align}
    &F_{\mathbf{I}}(s,c_{(\mathbf{I},\alpha)},q) = \frac{dc_{(\mathbf{I},\alpha)}}{dt} = \gamma_{(\mathbf{I},\alpha)}^{+} - \gamma_{(\mathbf{I},\alpha)}^{-} =  \sum_{i \ne \alpha} \left[\frac{c_{(\mathbf{I},i)}}{s} - \frac{c_{(\mathbf{I},\alpha)}}{s}\right], \nonumber\\ 
    &F_{\mathbf{C}}(s,c_{(\mathbf{C},\alpha)},q) = \frac{dc_{(\mathbf{C},\alpha)}}{dt} = \gamma_{(\mathbf{C},\alpha)}^{+} - \gamma_{(\mathbf{C},\alpha)}^{-} = \sum_{i \ne \alpha} \left[c_{(\mathbf{C},i)} c_{\alpha}^{q} - c_{(\mathbf{C},\alpha)} c_{i}^{q} \right].
    \label{eq:dcCdcIdt}
\end{align}
Similarly as in the annealed approach the system can reach the steady state in which all opinions are equinumerous or the one in which some states dominate over the others. Again, we can express all stationary states by $c$, which denotes the concentration of an arbitrarily chosen state, and by $c_{\mathbf{I}}$ and $c_{\mathbf{C}}$, which denote the concentration of independent and conformist agents in this state respectively:
\begin{align}
    &c_{0} = \dots = c_{s-(\xi + 1)} = c, \quad c_{s-\xi} = \dots = c_{s-1} = \frac{1-(s-\xi)c}{\xi}, \nonumber \\  
    &c_{(\mathbf{I},0)} = \dots = c_{(\mathbf{I},s-(\xi + 1))} = c_{\mathbf{I}}, \quad c_{(\mathbf{I},s-\xi)} = \dots = c_{(\mathbf{I},s-1)} = \frac{1-(s-\xi)c_{\mathbf{I}}}{\xi}, \nonumber \\ 
    &c_{(\mathbf{C},0)} = \dots = c_{(\mathbf{C},s-(\xi + 1))} = c_{\mathbf{C}}, \quad c_{(\mathbf{C},s-\xi)} = \dots = c_{(\mathbf{C},s-1)} = \frac{1-(s-\xi)c_{\mathbf{C}}}{\xi}, \nonumber \\
    &c = pc_{\mathbf{I}} + (1-p)c_{\mathbf{C}}. 
    \label{eq:CCiCcxi}
\end{align}
where $\xi = 1,2,\dots,s-1$ and $\xi=0$ indicates solution, where all states are equinumerous.
 
\begin{figure}[ht!]
	\centering
	\includegraphics[width=\linewidth]{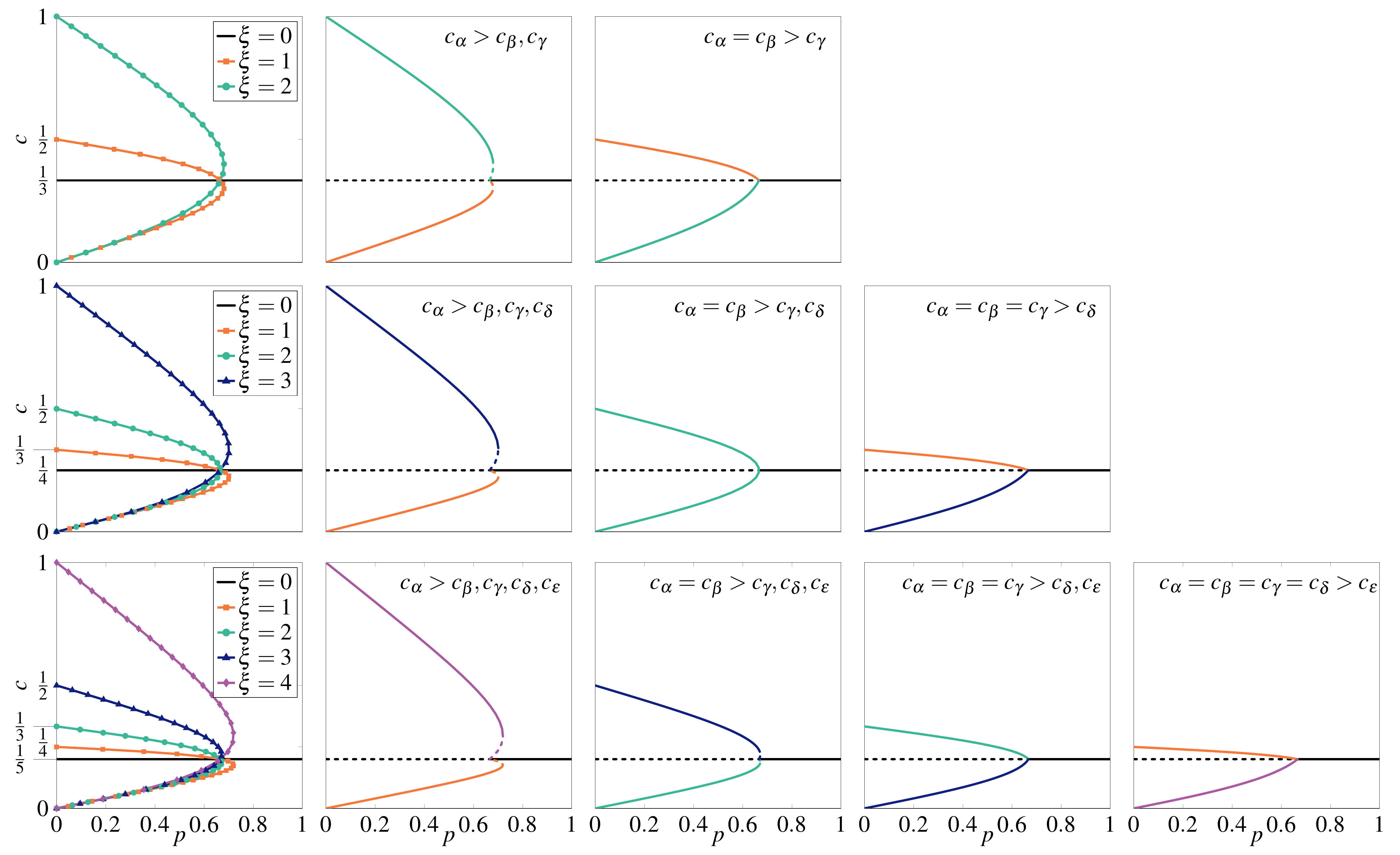}
	\caption{Steady states, given by the solution of Eq. (\ref{eq:pxiquenched}), for the quenched model with $q = 3$. The distribution of opinions is the same for Independent and Conformist agents. Each row corresponds to a different number of states: $s=3$ (top panels), $s=4$ (middle panels), $s=5$ (bottom panels). The first column represents all possible solutions indexed with different values of $\xi$ without the distinction between the stable and unstable solutions. The remaining columns represent stationary states for  initial conditions indicated in the top right corner of each panel, where $\alpha, \beta, \gamma, \delta, \epsilon \in \{0,1,\dots,s-1\}$. Stable solutions are denoted by the solid lines, whereas the unstable ones are marked with the dashed lines.}
	\label{fig:xieqquenched}
\end{figure}

Hence Eq. (\ref{eq:dcCdcIdt}) reduces to:
\begin{align}
    &F_{\mathbf{I}}(s,c_{\mathbf{I}},q,\xi) = \frac{1}{s}-c_{\mathbf{I}}, \nonumber\\ 
    &F_{\mathbf{C}}(s,c_{\mathbf{C}},q,\xi) = \xi \left[ \frac{1-(s-\xi)c_{\mathbf{C}}}{\xi} c^{q} - c_{\mathbf{C}} \left(\frac{1-(s-\xi)c}{\xi}\right)^{q} \right].
    \label{eq:FQxi}
\end{align}
Because in the steady state $F_{\mathbf{I}}(s,c_{\mathbf{I}_{st}},q,\xi) = 0$ and $F_{\mathbf{C}}(s,c_{\mathbf{C}_{st}},q,\xi) = 0$ we obtain: 
\begin{align}
    &c_{\mathbf{I}_{st}} = \frac{1}{s}, \nonumber\\ 
    &c_{\mathbf{C}_{st}} = \frac{c^{q}}{c^{q}(s-\xi) + \xi \left(\frac{1-(s-\xi)c}{\xi}\right)^{q}}.
    \label{eq:CiCcst}
\end{align}
By inserting the above formulas to the last formula of Eq. (\ref{eq:CCiCcxi}), we obtain:
\begin{equation}
    p = \frac{s \left[ c_{st}\xi \left(\frac{1-(s-\xi)c_{st}}{\xi}\right)^{q} - c_{st}^{q} \left[1-(s-\xi)c_{st}\right] \right]}{\xi \left(\frac{1-(s-\xi)c_{st}}{\xi}\right)^{q} - \xi c_{st}^{q}}.
    \label{eq:pxiquenched}
\end{equation}
It is easy to notice that the above equation for $s=2$ and $\xi = 1$ reproduces the analytical result for the quenched binary $q$-voter model \cite{Jed:Szn:17}. For more states, namely $s>2$, above relation produces $s-1$ stationary solutions, for $\xi = 1,2,\dots,s-1$ in the same way as for annealed model, see Fig. \ref{fig:xieqquenched}.

The stability of a steady point is given by determinant and trace of the Jacobian matrix at this point \cite{Li:etal:16,Jed:Szn:20}
\begin{equation}
    \mathbf{J}_{(c_{\mathbf{I}_{st}},c_{\mathbf{C}_{st}})} =
    \begin{bmatrix}
    \frac{\partial F_{\mathbf{I}}}{\partial c_{\mathbf{I}}} & \frac{\partial F_{\mathbf{I}}}{\partial c_{\mathbf{C}}} \\
    \frac{\partial F_{\mathbf{C}}}{\partial c_{\mathbf{I}}} & \frac{\partial F_{\mathbf{C}}}{\partial c_{\mathbf{C}}} 
    \end{bmatrix}_{(c_{\mathbf{I}},c_{\mathbf{C}}) = (c_{\mathbf{I}_{st}},c_{\mathbf{C}_{st}})},
\end{equation}
where
\begin{align}
    &\frac{\partial F_{\mathbf{I}}}{\partial c_{\mathbf{I}}} = -1,\\
    &\frac{\partial F_{\mathbf{I}}}{\partial c_{\mathbf{C}}} = 0,\\
    &\frac{\partial F_{\mathbf{C}}}{\partial c_{\mathbf{I}}} = qp\left[c^{q-1} - (s-\xi)c_{\mathbf{C}}c^{q-1} + (s-\xi)c_{\mathbf{C}}\left(\frac{1-(s-\xi)c}{\xi}\right)^{q-1}\right],\\
    &\frac{\partial F_{\mathbf{C}}}{\partial c_{\mathbf{C}}} = q(1-p)\left[c^{q-1} - (s-\xi)c_{\mathbf{C}}c^{q-1} + (s-\xi)c_{\mathbf{C}}\left(\frac{1-(s-\xi)c}{\xi}\right)^{q-1}\right] - (s-\xi)c^{q} - \xi\left(\frac{1-(s-\xi)c}{\xi}\right)^q.
\end{align}
The state is stable if $\det[\mathbf{J}_{(c_{\mathbf{I}_{st}},c_{\mathbf{C}_{st}})}] > 0$ and $\mathrm{tr}[\mathbf{J}_{(c_{\mathbf{I}_{st}},c_{\mathbf{C}_{st}})}] < 0$. For the steady state $(c_{\mathbf{I}_{st}},c_{\mathbf{C}_{st}}) = (\frac{1}{s},\frac{1}{s})$ we are able to determine the stability analytically as for the annealed version of the model:
\begin{align}
   &\det[\mathbf{J}_{\frac{1}{s},\frac{1}{s}}] = \left(\frac{1}{s}\right)^{q-1}(1-q+qp)\\
   &\mathrm{tr}[\mathbf{J}_{\frac{1}{s},\frac{1}{s}}] = \left(\frac{1}{s}\right)^{q-1}(q(1-p)-1) - 1.
\end{align}
Thus the stead state is stable for $p > p_{1}^{*}$ and unstable otherwise, where
\begin{equation}
    p_{1}^{*} = \frac{q-1}{q}.
    \label{eq:quenchedspinodal}
\end{equation}
We see that, the critical point $p_{1}^{*}$ depends only on the size of the group of influence $q$, but not on the number of states $s$, contrary to the annealed model. 

\begin{figure}[ht!]
	\centering
	\includegraphics[width=0.8\linewidth]{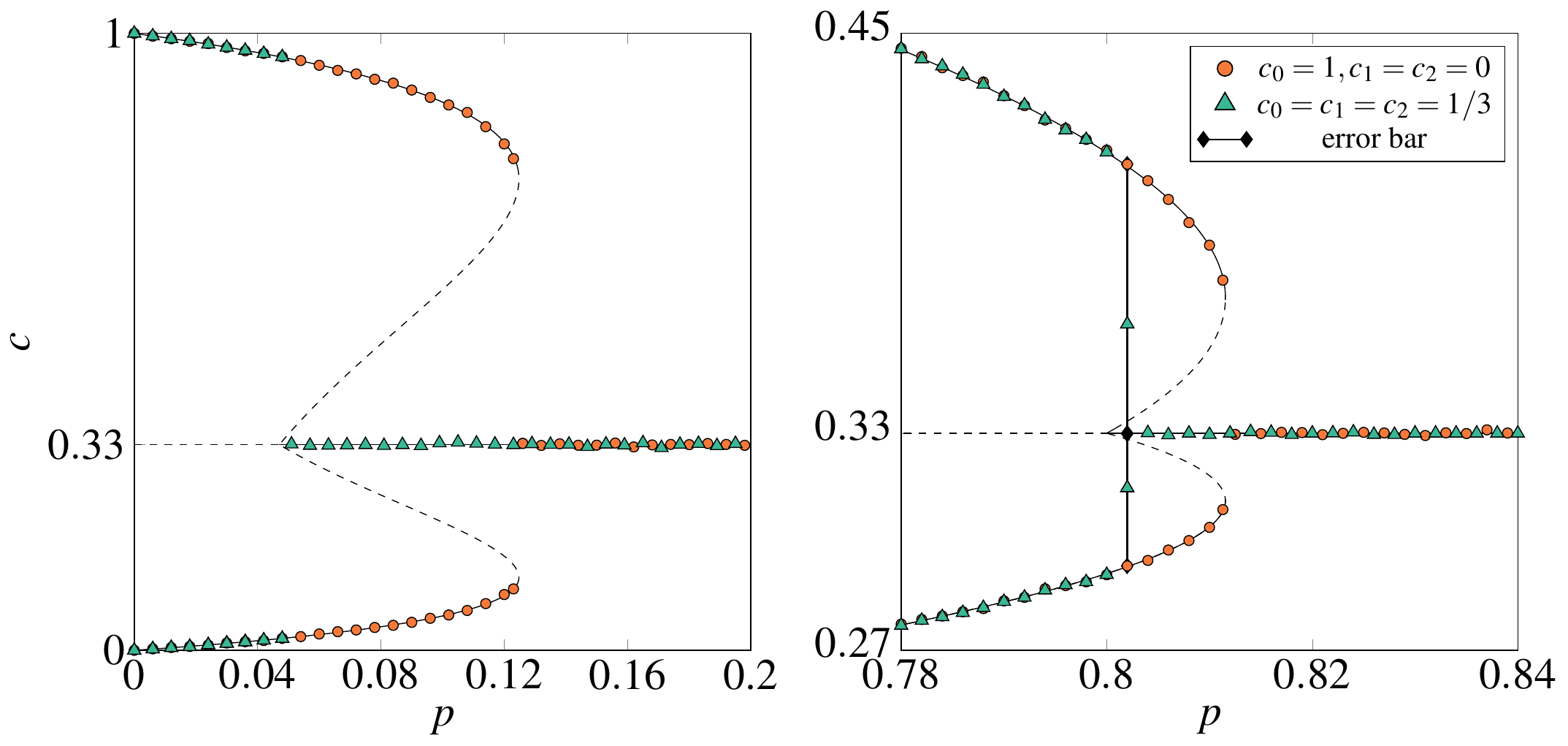}
	\caption{Dependence between the stationary concentration of agents in state $0$ and probability of independence $p$ within the annealed (left panel) and the quenched (right panel) approach for $q=5$ and $s=3$.  Lines represent  the solutions of Eq. (\ref{eq:pxiannealed}) and (\ref{eq:pxiquenched}): solid and dashed lines correspond to stable and unstable steady states, respectively. Symbols represent the outcome from MC simulations for the system size $N = 5 \cdot 10^{5}$. The results are averaged over 10 runs and collected after $t = 10^{5}$ MCS. Simulations are performed from two different initial conditions indicated in the legend. To compare analytical with MC results we plotted also error bars, but for almost all values of $p$ they are invisible, i.e. smaller than the symbols representing results.}
	\label{fig:hysteresis}
\end{figure}

\subsection{Discussion of the results}
\begin{figure}[ht!]
	\centering
	\includegraphics[width=\linewidth]{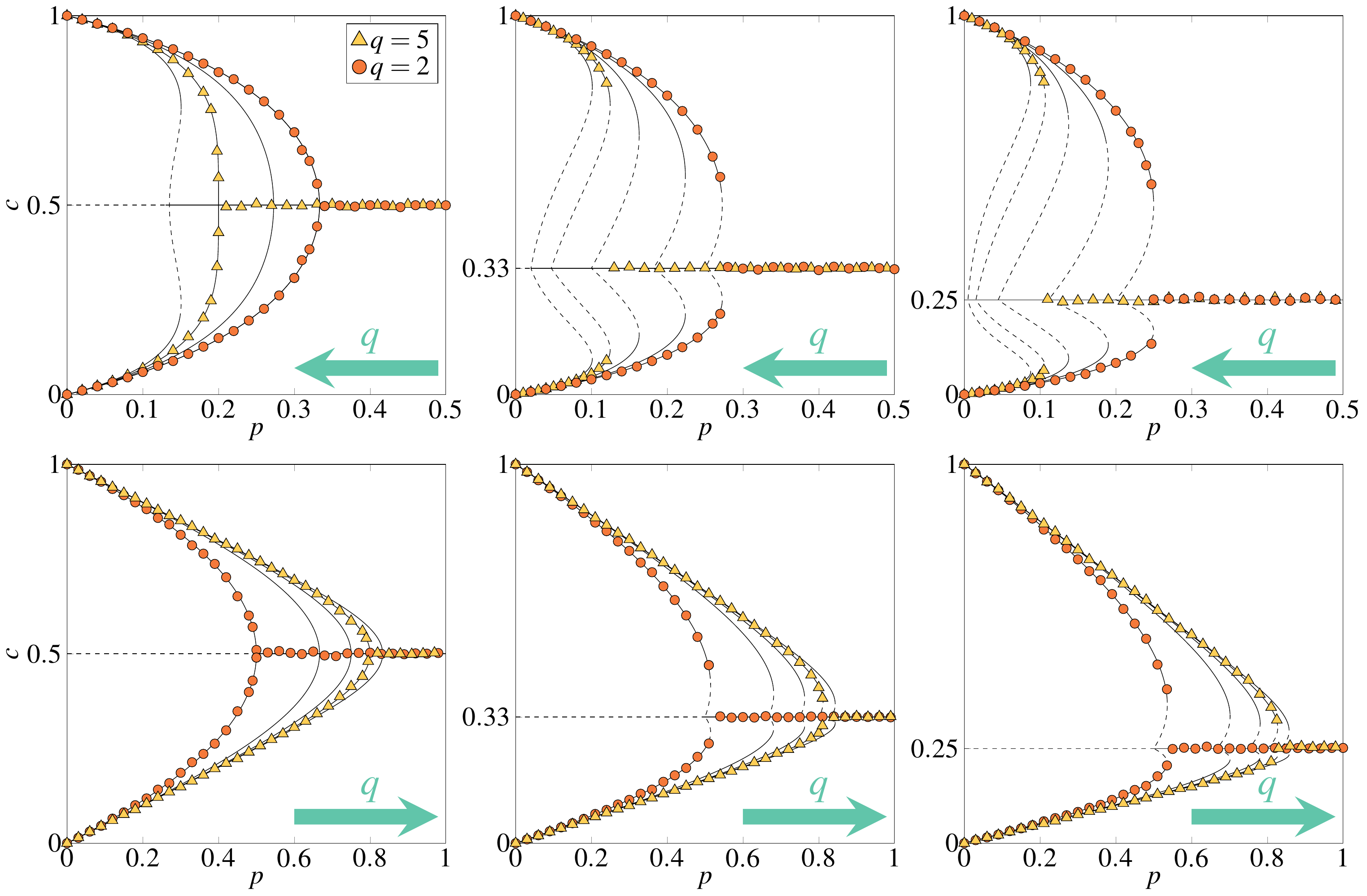}
	\caption{Dependence between the stationary concentration of agents in state $0$ and probability of independence $p$ within the annealed (upper panels) and quenched (bottom panels) approach for different values of the influence group size $q=\{2,3,4,5,6\}$. Arrows in the right corners of subplots indicate the direction in which $q$ increases. The number of states: $s = 2$ (left column), $s = 3$ (middle column) and  $s = 4$ (right column). Lines represent  the solutions of Eq. (\ref{eq:pxiannealed}) and (\ref{eq:pxiquenched}): solid and dashed lines correspond to stable and unstable steady states, respectively. Note that for $s=2$ we have only 4 curves, the reason for that is that for $q=2$ and $q=3$ exactly the same results are obtained. Symbols represent the outcome from MC simulations for the system size $N = 5 \cdot 10^{5}$ performed from initial condition $c_0 = 1, c_1=c_2=0$. The results are averaged over 10 runs and collected after $t = 2\cdot 10^{4}$ MCS. Symbols above the line $c=1/s$ correspond to the concentration of state 0, whereas symbols below the line $c=1/s$  represent concentration of all others. }
	\label{fig:codp}
\end{figure}

In the above sections, several aspects of the multi-state qVM was analyzed, namely the role of the parameters: $q$ being the size of the group of influence, $s$ being the number of states, as well as the type of the disorder. The model was considered on the complete graph, which allowed for the mean-field approach. However, all analytical results were also confirmed by the Monte Carlo simulations. In particular, we observe very good agreement between Eqs. (\ref{eq:pxiannealed}), (\ref{eq:annealedspinodal}), (\ref{eq:pxiquenched}), (\ref{eq:quenchedspinodal}) and numerical results for the critical points, see Figs. \ref{fig:hysteresis} and \ref{fig:codp}.

It was shown previously that under the quenched disorder only continuous phase transitions are possible within the original (binary) $q$-voter model with noise \cite{Jed:Szn:17}. Moreover, even under the annealed approach, the appropriate size of the influence group $q>5$ is required to obtain discontinuous phase transition \cite{Nyc:Szn:Cis:12, Per:etal:18, Vie:etal:20}. 

Here we have shown that already for the 3-state opinions, the model displays discontinuous phase transitions for any $q>1$, as presented in Fig. \ref{fig:codp}. An analogous result was obtained for the majority-vote model, in which agents are not influenced by the unanimous group of $q$ neighbors but by the absolute majority of all agents in the neighborhood. Within such a model with binary opinions, only continuous phase transitions appear \cite{Oli:92}, even if an additional noise is introduced \cite{Enc:etal:19, Vie:Cro:16, Vil:Mor:09}. However, for more than two states the majority-vote model displays discontinuous order-disorder phase transitions \cite{Li:etal:16,Che:Li:18}.

In Fig. \ref{fig:codp} it is also seen that discontinuous phase transitions are observed even under the quenched disorder if only the number of states is larger than two, although indeed they are less sharp. This result cannot be compared directly with the analogous one for the majority-vote model, because to our best knowledge multi-state majority-vote model was not studied with the quenched noise. However, the 3-state majority-vote model was studied on the quenched networks and in such a case only a continuous phase transitions were observed as in the binary model \cite{Mel:etal:10,Li:etal:16}.

\begin{figure}[ht!]
	\centering
	\includegraphics[width=\linewidth]{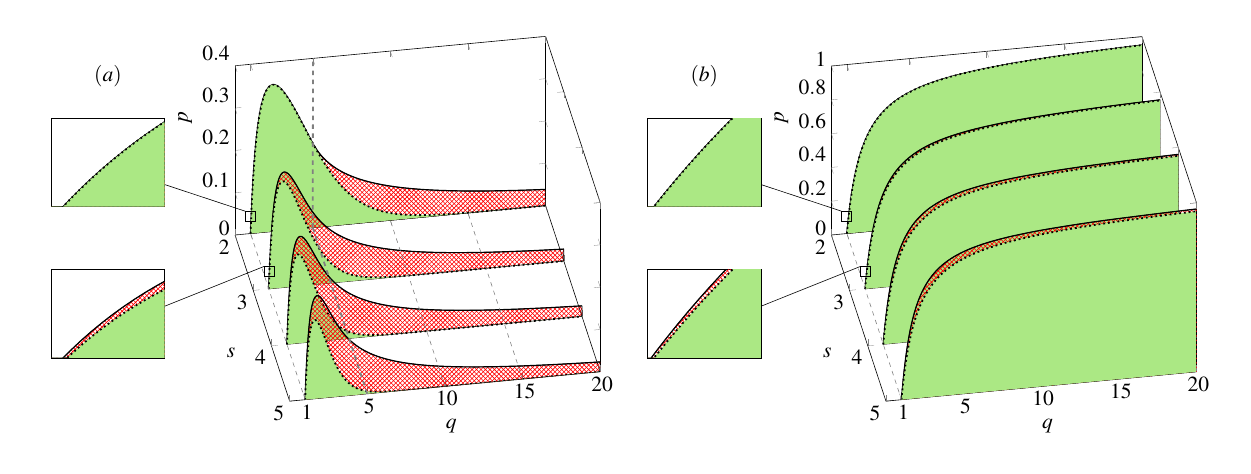}
	\caption{Phase diagram of the multi-state $q$-voter model under $(a)$ the annealed and $(b)$ the quenched approach. The ordered phases are marked by solid fill-color (green). The coexistence regions are marked by crosshatched pattern (red). The disordered phases are shown as no-fill-color regions (white). Lower and upper spinodals are marked by dotted and solid thick line respectively.}
	\label{fig:spinodals}
\end{figure}

Although discontinuous phase transitions are observed under both types of disorder, there is a huge difference between two approaches, clearly seen in Figs. \ref{fig:hysteresis}, \ref{fig:codp} and \ref{fig:spinodals}:
\begin{enumerate} 
\item For an arbitrary number of states $s$, spinodals $p_{1}^*$ and $p_{2}^*$ are non-monotonic functions of $q$ within the annealed approach (left panel in Fig. \ref{fig:spinodals}), whereas monotonically increasing ones under the quenched approach. 
\item While the parameter $q$ affects the lower spinodal $p_{1}^*$ under both approaches (differently as stayed above), parameter $s$ influences $p_{1}^*$ only in the case of the annealed approach, see Eq. (\ref{eq:annealedspinodal}) for the annealed approach and Eq. (\ref{eq:quenchedspinodal}) for the quenched one. 
\item Hysteresis, and simultaneously coexistence phase, appears under both approaches for $s>2$ but it is much larger under the annealed approach than under the quenched one. 
\end{enumerate}

\section{Conclusions}
Binary opinions are probably the most frequently used microscopic dynamical variables in models of opinion dynamics, such as the linear 
voter \cite{Red:19,Hol:Lig:75,Cas:For:Lor:09,Mob:03,Per:Kha:Tor:20} 
and non-linear voter \cite{Nyc:Szn:Cis:12,Abr:Paw:Szn:19,Jed:Szn:17,Per:etal:18,Vie:etal:20,Cas:Mun:Pas:09,Jed:Szn:20,Mor:etal:13,Gra:Kra:20,Muk:Maz:Roy:20,Tan:Mas:13} models, 
or the 
majority-vote model \cite{Vie:Cro:16,Che:etal:17,Enc:etal:18,Enc:etal:19,Vil:Mor:09,Oli:92,Kra:20,Kra:18,Kra:17,Vil:etal:19,Kra:Red:03}. However, it seems that the binary opinion format is not always sufficient and thus the multi-state versions of the 
voter \cite{Red:19,Kha:Gall:20,Vaz:Los:Bag:19,Boh:Gro:12,Her:Gal:19,Sta:Bar:Sat:12}, as well as 
majority-vote model \cite{Li:etal:16,Oes:Pir:Cro:19,Che:Red:05,Mel:etal:10,Vil:etal:20,Che:Li:18} 
was introduced.

In this paper we proposed the generalized version of the noisy  $q$-voter model, in which agents are described by the $s$-state dynamical variables. In our model all opinions are equivalent and agents can switch between any of them. Hence, it is not the best model for opinions that can be measured within the Likert psychometric scale,  used to scaling responses in survey research. Such a scale is often used to measure the level of agreement/disagreement, e.g., a typical five-level scale would be: Strongly disagree, Disagree, Neither agree nor disagree, Agree, Strongly agree. One may argue that going in one step from one extreme to another would be not very realistic. Therefore, the multi-state model introduced here would me more appropriate for making a choice between equivalent items. A good example of such a situation is a choice between equivalent products  or services  on the oligopoly market, such as the choice of the Cable Television and Cellular Phone Services or Automobiles. The model, which could describe opinions on the Likert scale requires in our opinion additional assumptions, such as bounded confidence, and will be studied in the future.

We have investigated the model under two types of approaches, the annealed and the quenched one, to check how the type of disorder influences the model for $s>2$. Previously it was shown that for $s=2$ quenched disorder forbids discontinuous phase transitions \cite{Jed:Szn:17}. However, it occurs that for $s>2$ discontinuous phase transitions are possible even for the quenched disorder. Moreover, they appear for any $q>1$, on contrary to the original binary $q$-voter model for which discontinuous phase transitions appear only for $q>5$ within the annealed approach.

Physicists always look for universalities and this is also the case in this paper. If we compare two popular, yet very different, binary models of opinion dynamics, such as the majority-vote and the $q$-voter model we clearly see such a universality. In both models introducing only one additional (third) state results in discontinuous phase transitions for the annealed approach. The universality of the second result obtained here, namely the survival of the discontinuous phase transition under the quenched approach would be an interesting task for the future.
  

\begin{thebibliography}{10}
\urlstyle{rm}
\expandafter\ifx\csname url\endcsname\relax
  \def\url#1{\texttt{#1}}\fi
\expandafter\ifx\csname urlprefix\endcsname\relax\def\urlprefix{URL }\fi
\expandafter\ifx\csname doiprefix\endcsname\relax\def\doiprefix{DOI: }\fi
\providecommand{\bibinfo}[2]{#2}
\providecommand{\eprint}[2][]{\url{#2}}

\bibitem{Sche:Wes:Bro:03}
\bibinfo{author}{Scheffer, M.}, \bibinfo{author}{Westley, F.} \&
  \bibinfo{author}{Brock, W.}
\newblock \bibinfo{journal}{\bibinfo{title}{Slow response of societies to new
  problems: Causes and costs}}.
\newblock {\emph{\JournalTitle{Ecosystems}}} \textbf{\bibinfo{volume}{6}},
  \bibinfo{pages}{493--502}, \doiprefix\url{10.1007/s10021-002-0146-0}
  (\bibinfo{year}{2003}).

\bibitem{Bis:15}
\bibinfo{author}{Bissell, J.}, \bibinfo{author}{Caiado, C.},
  \bibinfo{author}{Curtis, S.}, \bibinfo{author}{Goldstein, M.} \&
  \bibinfo{author}{Straughan, B.}
\newblock \emph{\bibinfo{title}{Tipping Points: Modelling Social Problems and
  Health}} (\bibinfo{publisher}{Wiley}, \bibinfo{year}{2015}).

\bibitem{Str:Liz:17}
\bibinfo{author}{Strand, M.} \& \bibinfo{author}{Lizardo, O.}
\newblock \bibinfo{journal}{\bibinfo{title}{The hysteresis effect: Theorizing
  mismatch in action}}.
\newblock {\emph{\JournalTitle{Journal for the Theory of Social Behaviour}}}
  \textbf{\bibinfo{volume}{47}}, \bibinfo{pages}{164--194},
  \doiprefix\url{10.1111/jtsb.12117} (\bibinfo{year}{2017}).

\bibitem{Pru:etal:18}
\bibinfo{author}{Pruitt, J.} \emph{et~al.}
\newblock \bibinfo{journal}{\bibinfo{title}{Social tipping points in animal
  societies}}.
\newblock {\emph{\JournalTitle{Proceedings of the Royal Society B: Biological
  Sciences}}} \textbf{\bibinfo{volume}{285}},
  \doiprefix\url{10.1098/rspb.2018.1282} (\bibinfo{year}{2018}).

\bibitem{Cen:etal:18}
\bibinfo{author}{Centola, D.}, \bibinfo{author}{Becker, J.},
  \bibinfo{author}{Brackbill, D.} \& \bibinfo{author}{Baronchelli, A.}
\newblock \bibinfo{journal}{\bibinfo{title}{Experimental evidence for tipping
  points in social convention}}.
\newblock {\emph{\JournalTitle{Science}}} \textbf{\bibinfo{volume}{360}},
  \bibinfo{pages}{1116--1119}, \doiprefix\url{10.1126/science.aas8827}
  (\bibinfo{year}{2018}).

\bibitem{Nyc:Szn:Cis:12}
\bibinfo{author}{Nyczka, P.}, \bibinfo{author}{Sznajd-Weron, K.} \&
  \bibinfo{author}{Cis\l{}o, J.}
\newblock \bibinfo{journal}{\bibinfo{title}{Phase transitions in the $q$-voter
  model with two types of stochastic driving}}.
\newblock {\emph{\JournalTitle{Phys. Rev. E}}} \textbf{\bibinfo{volume}{86}},
  \bibinfo{pages}{011105}, \doiprefix\url{10.1103/PhysRevE.86.011105}
  (\bibinfo{year}{2012}).

\bibitem{Vie:Cro:16}
\bibinfo{author}{Vieira, A.} \& \bibinfo{author}{Crokidakis, N.}
\newblock \bibinfo{journal}{\bibinfo{title}{Phase transitions in the
  majority-vote model with two types of noises}}.
\newblock {\emph{\JournalTitle{Physica A: Statistical Mechanics and its
  Applications}}} \textbf{\bibinfo{volume}{450}}, \bibinfo{pages}{30--36},
  \doiprefix\url{10.1016/j.physa.2016.01.013} (\bibinfo{year}{2016}).

\bibitem{Che:etal:17}
\bibinfo{author}{Chen, H.} \emph{et~al.}
\newblock \bibinfo{journal}{\bibinfo{title}{First-order phase transition in a
  majority-vote model with inertia}}.
\newblock {\emph{\JournalTitle{Physical Review E}}}
  \textbf{\bibinfo{volume}{95}}, \doiprefix\url{10.1103/PhysRevE.95.042304}
  (\bibinfo{year}{2017}).

\bibitem{Tuz:Fer:Equ:18}
\bibinfo{author}{Tuzón, P.}, \bibinfo{author}{Fernández-Gracia, J.} \&
  \bibinfo{author}{Eguíluz, V.}
\newblock \bibinfo{journal}{\bibinfo{title}{From continuous to discontinuous
  transitions in social diffusion}}.
\newblock {\emph{\JournalTitle{Frontiers in Physics}}}
  \textbf{\bibinfo{volume}{6}}, \doiprefix\url{10.3389/fphy.2018.00021}
  (\bibinfo{year}{2018}).

\bibitem{Enc:etal:18}
\bibinfo{author}{Encinas, J.}, \bibinfo{author}{Harunari, P.},
  \bibinfo{author}{De~Oliveira, M.} \& \bibinfo{author}{Fiore, C.}
\newblock \bibinfo{journal}{\bibinfo{title}{Fundamental ingredients for
  discontinuous phase transitions in the inertial majority vote model}}.
\newblock {\emph{\JournalTitle{Scientific Reports}}}
  \textbf{\bibinfo{volume}{8}}, \doiprefix\url{10.1038/s41598-018-27240-4}
  (\bibinfo{year}{2018}).

\bibitem{Enc:etal:19}
\bibinfo{author}{Encinas, J.}, \bibinfo{author}{Chen, H.},
  \bibinfo{author}{de~Oliveira, M.} \& \bibinfo{author}{Fiore, C.}
\newblock \bibinfo{journal}{\bibinfo{title}{Majority vote model with ancillary
  noise in complex networks}}.
\newblock {\emph{\JournalTitle{Physica A: Statistical Mechanics and its
  Applications}}} \textbf{\bibinfo{volume}{516}}, \bibinfo{pages}{563--570},
  \doiprefix\url{10.1016/j.physa.2018.10.055} (\bibinfo{year}{2019}).

\bibitem{Now:Szn:19}
\bibinfo{author}{Nowak, B.} \& \bibinfo{author}{Sznajd-Weron, K.}
\newblock \bibinfo{journal}{\bibinfo{title}{Homogeneous symmetrical threshold
  model with nonconformity: Independence versus anticonformity,}}.
\newblock {\emph{\JournalTitle{Complexity}}} \textbf{\bibinfo{volume}{2019}},
  \doiprefix\url{10.1155/2019/5150825} (\bibinfo{year}{2019}).

\bibitem{Abr:Paw:Szn:19}
\bibinfo{author}{Abramiuk, A.}, \bibinfo{author}{Pawłowski, J.} \&
  \bibinfo{author}{Sznajd-Weron, K.}
\newblock \bibinfo{journal}{\bibinfo{title}{Is independence necessary for a
  discontinuous phase transition within the q-voter model?}}
\newblock {\emph{\JournalTitle{Entropy}}} \textbf{\bibinfo{volume}{21}},
  \doiprefix\url{10.3390/e21050521} (\bibinfo{year}{2019}).

\bibitem{Chm:etal:20}
\bibinfo{author}{Chmiel, A.}, \bibinfo{author}{Sienkiewicz, J.},
  \bibinfo{author}{Fronczak, A.} \& \bibinfo{author}{Fronczak, P.}
\newblock \bibinfo{journal}{\bibinfo{title}{A veritable zoology of successive
  phase transitions in the asymmetric q-voter model on multiplex networks}}.
\newblock {\emph{\JournalTitle{Entropy}}} \textbf{\bibinfo{volume}{22}},
  \doiprefix\url{10.3390/e22091018} (\bibinfo{year}{2020}).

\bibitem{Aiz:Weh:89}
\bibinfo{author}{Aizenman, M.} \& \bibinfo{author}{Wehr, J.}
\newblock \bibinfo{journal}{\bibinfo{title}{Rounding of first-order phase
  transitions in systems with quenched disorder}}.
\newblock {\emph{\JournalTitle{Physical Review Letters}}}
  \textbf{\bibinfo{volume}{62}}, \bibinfo{pages}{2503--2506},
  \doiprefix\url{10.1103/PhysRevLett.62.2503} (\bibinfo{year}{1989}).

\bibitem{Bor:Mar:Mun:13}
\bibinfo{author}{Borile, C.}, \bibinfo{author}{Maritan, A.} \&
  \bibinfo{author}{Muñoz, M.}
\newblock \bibinfo{journal}{\bibinfo{title}{The effect of quenched disorder in
  neutral theories}}.
\newblock {\emph{\JournalTitle{Journal of Statistical Mechanics: Theory and
  Experiment}}} \textbf{\bibinfo{volume}{2013}},
  \doiprefix\url{10.1088/1742-5468/2013/04/P04032} (\bibinfo{year}{2013}).

\bibitem{Vil:Bon:Mun:14}
\bibinfo{author}{Villa~Martín, P.}, \bibinfo{author}{Bonachela, J.} \&
  \bibinfo{author}{Muñoz, M.}
\newblock \bibinfo{journal}{\bibinfo{title}{Quenched disorder forbids
  discontinuous transitions in nonequilibrium low-dimensional systems}}.
\newblock {\emph{\JournalTitle{Physical Review E - Statistical, Nonlinear, and
  Soft Matter Physics}}} \textbf{\bibinfo{volume}{89}},
  \doiprefix\url{10.1103/PhysRevE.89.012145} (\bibinfo{year}{2014}).

\bibitem{Jed:Szn:17}
\bibinfo{author}{J{\k{e}}drzejewski, A.} \& \bibinfo{author}{Sznajd-Weron, K.}
\newblock \bibinfo{journal}{\bibinfo{title}{Person-situation debate revisited:
  Phase transitions with quenched and annealed disorders}}.
\newblock {\emph{\JournalTitle{Entropy}}} \textbf{\bibinfo{volume}{19}},
  \bibinfo{pages}{415}, \doiprefix\url{10.3390/e19080415}
  (\bibinfo{year}{2017}).

\bibitem{Wu:82}
\bibinfo{author}{Wu, F.~Y.}
\newblock \bibinfo{journal}{\bibinfo{title}{The potts model}}.
\newblock {\emph{\JournalTitle{Reviews of Modern Physics}}}
  \textbf{\bibinfo{volume}{54}}, \bibinfo{pages}{253--268}
  (\bibinfo{year}{1982}).

\bibitem{Vil:Mor:09}
\bibinfo{author}{Vilela, A. L.~M.} \& \bibinfo{author}{Moreira, F. G.~B.}
\newblock \bibinfo{journal}{\bibinfo{title}{Majority-vote model with different
  agents}}.
\newblock {\emph{\JournalTitle{Physica A: Statistical Mechanics and its
  Applications}}} \textbf{\bibinfo{volume}{388}}, \bibinfo{pages}{4171--4178},
  \doiprefix\url{10.1016/j.physa.2009.06.046} (\bibinfo{year}{2009}).

\bibitem{Li:etal:16}
\bibinfo{author}{Li, G.}, \bibinfo{author}{Chen, H.}, \bibinfo{author}{Huang,
  F.} \& \bibinfo{author}{Shen, C.}
\newblock \bibinfo{journal}{\bibinfo{title}{Discontinuous phase transition in
  an annealed multi-state majority-vote model}}.
\newblock {\emph{\JournalTitle{Journal of Statistical Mechanics: Theory and
  Experiment}}} \textbf{\bibinfo{volume}{2016}},
  \doiprefix\url{10.1088/1742-5468/2016/07/073403} (\bibinfo{year}{2016}).

\bibitem{Oes:Pir:Cro:19}
\bibinfo{author}{Oestereich, A.}, \bibinfo{author}{Pires, M.} \&
  \bibinfo{author}{Crokidakis, N.}
\newblock \bibinfo{journal}{\bibinfo{title}{Three-state opinion dynamics in
  modular networks}}.
\newblock {\emph{\JournalTitle{Physical Review E}}}
  \textbf{\bibinfo{volume}{100}}, \doiprefix\url{10.1103/PhysRevE.100.032312}
  (\bibinfo{year}{2019}).

\bibitem{Red:19}
\bibinfo{author}{Redner, S.}
\newblock \bibinfo{journal}{\bibinfo{title}{Reality-inspired voter models: A
  mini-review}}.
\newblock {\emph{\JournalTitle{Comptes Rendus Physique}}}
  \textbf{\bibinfo{volume}{20}}, \bibinfo{pages}{275 -- 292},
  \doiprefix\url{https://doi.org/10.1016/j.crhy.2019.05.004}
  (\bibinfo{year}{2019}).

\bibitem{Kha:Gall:20}
\bibinfo{author}{N., K.} \& \bibinfo{author}{T., G.}
\newblock \bibinfo{title}{Zealots in multi-state noisy voter models}
  (\bibinfo{year}{2020}).
\newblock \bibinfo{note}{ArXiv:2007.07535}.

\bibitem{Vaz:Los:Bag:19}
\bibinfo{author}{Vazquez, F.}, \bibinfo{author}{Loscar, E.~S.} \&
  \bibinfo{author}{Baglietto, G.}
\newblock \bibinfo{journal}{\bibinfo{title}{Multistate voter model with
  imperfect copying}}.
\newblock {\emph{\JournalTitle{Phys. Rev. E}}} \textbf{\bibinfo{volume}{100}},
  \bibinfo{pages}{042301}, \doiprefix\url{10.1103/PhysRevE.100.042301}
  (\bibinfo{year}{2019}).

\bibitem{Boh:Gro:12}
\bibinfo{author}{Böhme, G.} \& \bibinfo{author}{Gross, T.}
\newblock \bibinfo{journal}{\bibinfo{title}{Fragmentation transitions in
  multistate voter models}}.
\newblock {\emph{\JournalTitle{Physical Review E - Statistical, Nonlinear, and
  Soft Matter Physics}}} \textbf{\bibinfo{volume}{85}},
  \doiprefix\url{10.1103/PhysRevE.85.066117} (\bibinfo{year}{2012}).

\bibitem{Her:Gal:19}
\bibinfo{author}{Herreriás-Azcué, F.} \& \bibinfo{author}{Galla, T.}
\newblock \bibinfo{journal}{\bibinfo{title}{Consensus and diversity in
  multistate noisy voter models}}.
\newblock {\emph{\JournalTitle{Physical Review E}}}
  \textbf{\bibinfo{volume}{100}}, \doiprefix\url{10.1103/PhysRevE.100.022304}
  (\bibinfo{year}{2019}).

\bibitem{Sta:Bar:Sat:12}
\bibinfo{author}{Starnini, M.}, \bibinfo{author}{Baronchelli, A.} \&
  \bibinfo{author}{Pastor-Satorras, R.}
\newblock \bibinfo{journal}{\bibinfo{title}{Ordering dynamics of the
  multi-state voter model}}.
\newblock {\emph{\JournalTitle{Journal of Statistical Mechanics: Theory and
  Experiment}}} \textbf{\bibinfo{volume}{2012}}, \bibinfo{pages}{P10027},
  \doiprefix\url{10.1088/1742-5468/2012/10/p10027} (\bibinfo{year}{2012}).

\bibitem{Che:Red:05}
\bibinfo{author}{Chen, P.} \& \bibinfo{author}{Redner, S.}
\newblock \bibinfo{journal}{\bibinfo{title}{Consensus formation in multi-state
  majority and plurality models}}.
\newblock {\emph{\JournalTitle{Journal of Physics A: Mathematical and
  General}}} \textbf{\bibinfo{volume}{38}}, \bibinfo{pages}{7239--7252},
  \doiprefix\url{10.1088/0305-4470/38/33/003} (\bibinfo{year}{2005}).

\bibitem{Mel:etal:10}
\bibinfo{author}{Melo, D.}, \bibinfo{author}{Pereira, L.} \&
  \bibinfo{author}{Moreira, F.}
\newblock \bibinfo{journal}{\bibinfo{title}{The phase diagram and critical
  behavior of the three-state majority-vote model}}.
\newblock {\emph{\JournalTitle{Journal of Statistical Mechanics: Theory and
  Experiment}}} \textbf{\bibinfo{volume}{2010}},
  \doiprefix\url{10.1088/1742-5468/2010/11/P11032} (\bibinfo{year}{2010}).

\bibitem{Vil:etal:20}
\bibinfo{author}{Vilela, A.} \emph{et~al.}
\newblock \bibinfo{journal}{\bibinfo{title}{Three-state majority-vote model on
  scale-free networks and the unitary relation for critical exponents}}.
\newblock {\emph{\JournalTitle{Scientific Reports}}}
  \textbf{\bibinfo{volume}{10}}, \doiprefix\url{10.1038/s41598-020-63929-1}
  (\bibinfo{year}{2020}).

\bibitem{Che:Li:18}
\bibinfo{author}{Chen, H.} \& \bibinfo{author}{Li, G.}
\newblock \bibinfo{journal}{\bibinfo{title}{Phase transitions in a multistate
  majority-vote model on complex networks}}.
\newblock {\emph{\JournalTitle{Physical Review E}}}
  \textbf{\bibinfo{volume}{97}}, \doiprefix\url{10.1103/PhysRevE.97.062304}
  (\bibinfo{year}{2018}).

\bibitem{Ban:Mal:19}
\bibinfo{author}{Bańcerowski, P.} \& \bibinfo{author}{Malarz, K.}
\newblock \bibinfo{journal}{\bibinfo{title}{Multi-choice opinion dynamics model
  based on latané theory}}.
\newblock {\emph{\JournalTitle{European Physical Journal B}}}
  \textbf{\bibinfo{volume}{92}}, \doiprefix\url{10.1140/epjb/e2019-90533-0}
  (\bibinfo{year}{2019}).

\bibitem{SznWer:Szn:Wer:20}
\bibinfo{author}{Sznajd-Weron, K.}, \bibinfo{author}{Sznajd, J.} \&
  \bibinfo{author}{Weron, T.}
\newblock \bibinfo{journal}{\bibinfo{title}{A review on the sznajd model — 20
  years after}}.
\newblock {\emph{\JournalTitle{Physica A: Statistical Mechanics and its
  Applications}}} \textbf{\bibinfo{volume}{565}}, \bibinfo{pages}{125537},
  \doiprefix\url{https://doi.org/10.1016/j.physa.2020.125537}
  (\bibinfo{year}{2021}).

\bibitem{Net:Bri:20}
\bibinfo{author}{Neto, M.} \& \bibinfo{author}{Brigatti, E.}
\newblock \bibinfo{journal}{\bibinfo{title}{Discontinuous transitions can
  survive to quenched disorder in a two-dimensional nonequilibrium system}}.
\newblock {\emph{\JournalTitle{Physical Review E}}}
  \textbf{\bibinfo{volume}{101}}, \doiprefix\url{10.1103/PhysRevE.101.022112}
  (\bibinfo{year}{2020}).

\bibitem{Per:etal:18}
\bibinfo{author}{Peralta, A.}, \bibinfo{author}{Carro, A.},
  \bibinfo{author}{San~Miguel, M.} \& \bibinfo{author}{R, T.}
\newblock \bibinfo{journal}{\bibinfo{title}{Analytical and numerical study of
  the non-linear noisy voter model on complex networks}}.
\newblock {\emph{\JournalTitle{Chaos}}} \textbf{\bibinfo{volume}{28}},
  \bibinfo{pages}{075516}, \doiprefix\url{10.1063/1.5030112}
  (\bibinfo{year}{2018}).

\bibitem{Vie:etal:20}
\bibinfo{author}{Vieira, A.}, \bibinfo{author}{Peralta, A.},
  \bibinfo{author}{Toral, R.}, \bibinfo{author}{Miguel, M.} \&
  \bibinfo{author}{Anteneodo, C.}
\newblock \bibinfo{journal}{\bibinfo{title}{Pair approximation for the noisy
  threshold q-voter model}}.
\newblock {\emph{\JournalTitle{Physical Review E}}}
  \textbf{\bibinfo{volume}{101}}, \doiprefix\url{10.1103/PhysRevE.101.052131}
  (\bibinfo{year}{2020}).

\bibitem{Cas:Mun:Pas:09}
\bibinfo{author}{Castellano, C.}, \bibinfo{author}{Mu{\~n}oz, M.~A.} \&
  \bibinfo{author}{Pastor-Satorras, R.}
\newblock \bibinfo{journal}{\bibinfo{title}{Nonlinear $q$-voter model}}.
\newblock {\emph{\JournalTitle{Physical Review E}}}
  \textbf{\bibinfo{volume}{80}}, \bibinfo{pages}{041129},
  \doiprefix\url{10.1103/PhysRevE.80.041129} (\bibinfo{year}{2009}).

\bibitem{Jed:Szn:20}
\bibinfo{author}{J\k{e}drzejewski, A.} \& \bibinfo{author}{Sznajd-Weron, K.}
\newblock \bibinfo{journal}{\bibinfo{title}{Nonlinear q -voter model from the
  quenched perspective}}.
\newblock {\emph{\JournalTitle{Chaos}}} \textbf{\bibinfo{volume}{30}},
  \doiprefix\url{10.1063/1.5134684} (\bibinfo{year}{2020}).

\bibitem{Mor:etal:13}
\bibinfo{author}{Moretti, P.}, \bibinfo{author}{Liu, S.},
  \bibinfo{author}{Castellano, C.} \& \bibinfo{author}{Pastor-Satorras, R.}
\newblock \bibinfo{journal}{\bibinfo{title}{Mean-field analysis of the
  $q$-voter model on networks}}.
\newblock {\emph{\JournalTitle{Journal of Statistical Physics}}}
  \textbf{\bibinfo{volume}{151}}, \bibinfo{pages}{113--130},
  \doiprefix\url{10.1007/s10955-013-0704-1} (\bibinfo{year}{2013}).

\bibitem{Str:94}
\bibinfo{author}{Strogatz, S.}
\newblock \emph{\bibinfo{title}{Nonlinear dynamics and chaos: with applications
  to physics, biology, chemistry, and engineering}}
  (\bibinfo{publisher}{Perseus Books Publishing}, \bibinfo{year}{1994}).

\bibitem{Now:Szn:20}
\bibinfo{author}{Nowak, B.} \& \bibinfo{author}{Sznajd-Weron, K.}
\newblock \bibinfo{journal}{\bibinfo{title}{Symmetrical threshold model with
  independence on random graphs}}.
\newblock {\emph{\JournalTitle{Physical Review E}}}
  \textbf{\bibinfo{volume}{101}}, \doiprefix\url{10.1103/PhysRevE.101.052316}
  (\bibinfo{year}{2020}).

\bibitem{Oli:92}
\bibinfo{author}{de~Oliveira, M.}
\newblock \bibinfo{journal}{\bibinfo{title}{Isotropic majority-vote model on a
  square lattice}}.
\newblock {\emph{\JournalTitle{Journal of Statistical Physics}}}
  \textbf{\bibinfo{volume}{66}}, \bibinfo{pages}{273--281},
  \doiprefix\url{10.1007/BF01060069} (\bibinfo{year}{1992}).

\bibitem{Hol:Lig:75}
\bibinfo{author}{Holley, T.~M., Richard A.;~Liggett}.
\newblock \bibinfo{journal}{\bibinfo{title}{Ergodic theorems for weakly
  interacting infinite systems and the voter model}}.
\newblock {\emph{\JournalTitle{Annals of Probability}}}
  \textbf{\bibinfo{volume}{3}}, \doiprefix\url{10.1214/aop/1176996306}
  (\bibinfo{year}{1975}).

\bibitem{Cas:For:Lor:09}
\bibinfo{author}{Castellano, C.}, \bibinfo{author}{Fortunato, S.} \&
  \bibinfo{author}{Loreto, V.}
\newblock \bibinfo{journal}{\bibinfo{title}{Statistical physics of social
  dynamics}}.
\newblock {\emph{\JournalTitle{Reviews of Modern Physics}}}
  \textbf{\bibinfo{volume}{81}}, \bibinfo{pages}{591--646},
  \doiprefix\url{10.1103/RevModPhys.81.591} (\bibinfo{year}{2009}).

\bibitem{Mob:03}
\bibinfo{author}{Mobilia, M.}
\newblock \bibinfo{journal}{\bibinfo{title}{Does a single zealot affect an
  infinite group of voters?}}
\newblock {\emph{\JournalTitle{Phys. Rev. Lett.}}}
  \textbf{\bibinfo{volume}{91}}, \bibinfo{pages}{028701},
  \doiprefix\url{10.1103/PhysRevLett.91.028701} (\bibinfo{year}{2003}).

\bibitem{Per:Kha:Tor:20}
\bibinfo{author}{Peralta, A.}, \bibinfo{author}{Khalil, N.} \&
  \bibinfo{author}{Toral, R.}
\newblock \bibinfo{journal}{\bibinfo{title}{Ordering dynamics in the voter
  model with aging}}.
\newblock {\emph{\JournalTitle{Physica A: Statistical Mechanics and its
  Applications}}} \textbf{\bibinfo{volume}{552}},
  \doiprefix\url{10.1016/j.physa.2019.122475} (\bibinfo{year}{2020}).

\bibitem{Gra:Kra:20}
\bibinfo{author}{Gradowski, T.} \& \bibinfo{author}{Krawiecki, A.}
\newblock \bibinfo{journal}{\bibinfo{title}{Pair approximation for the q-voter
  model with independence on multiplex networks}}.
\newblock {\emph{\JournalTitle{Physical Review E}}}
  \textbf{\bibinfo{volume}{102}}, \doiprefix\url{10.1103/PhysRevE.102.022314}
  (\bibinfo{year}{2020}).

\bibitem{Muk:Maz:Roy:20}
\bibinfo{author}{Mukhopadhyay, A.}, \bibinfo{author}{Mazumdar, R.} \&
  \bibinfo{author}{Roy, R.}
\newblock \bibinfo{journal}{\bibinfo{title}{Voter and majority dynamics with
  biased and stubborn agents}}.
\newblock {\emph{\JournalTitle{Journal of Statistical Physics}}}
  \textbf{\bibinfo{volume}{181}}, \bibinfo{pages}{1239--1265},
  \doiprefix\url{10.1007/s10955-020-02625-w} (\bibinfo{year}{2020}).

\bibitem{Tan:Mas:13}
\bibinfo{author}{Tanabe, S.} \& \bibinfo{author}{Masuda, N.}
\newblock \bibinfo{journal}{\bibinfo{title}{Complex dynamics of a nonlinear
  voter model with contrarian agents}}.
\newblock {\emph{\JournalTitle{Chaos: An Interdisciplinary Journal of Nonlinear
  Science}}} \textbf{\bibinfo{volume}{23}}, \bibinfo{pages}{043136},
  \doiprefix\url{10.1063/1.4851175} (\bibinfo{year}{2013}).

\bibitem{Kra:20}
\bibinfo{author}{Krawiecki, A.}
\newblock \bibinfo{journal}{\bibinfo{title}{Ferromagnetic and spin-glass-like
  transition in the majority vote model on complete and random graphs}}.
\newblock {\emph{\JournalTitle{European Physical Journal B}}}
  \textbf{\bibinfo{volume}{93}}, \doiprefix\url{10.1140/epjb/e2020-10288-9}
  (\bibinfo{year}{2020}).

\bibitem{Kra:18}
\bibinfo{author}{Krawiecki, A.}
\newblock \bibinfo{journal}{\bibinfo{title}{Spin-glass-like transition in the
  majority-vote model with anticonformists}}.
\newblock {\emph{\JournalTitle{European Physical Journal B}}}
  \textbf{\bibinfo{volume}{91}}, \doiprefix\url{10.1140/epjb/e2018-80551-9}
  (\bibinfo{year}{2018}).

\bibitem{Kra:17}
\bibinfo{author}{Krawiecki, A.}
\newblock \bibinfo{journal}{\bibinfo{title}{Stochastic resonance in the
  majority vote model on regular and small-world lattices}}.
\newblock {\emph{\JournalTitle{International Journal of Modern Physics B}}}
  \textbf{\bibinfo{volume}{31}}, \doiprefix\url{10.1142/S0217979217502149}
  (\bibinfo{year}{2017}).

\bibitem{Vil:etal:19}
\bibinfo{author}{Vilela, A.~L.}, \bibinfo{author}{Wang, C.},
  \bibinfo{author}{Nelson, K.~P.} \& \bibinfo{author}{Stanley, H.~E.}
\newblock \bibinfo{journal}{\bibinfo{title}{Majority-vote model for financial
  markets}}.
\newblock {\emph{\JournalTitle{Physica A: Statistical Mechanics and its
  Applications}}} \textbf{\bibinfo{volume}{515}}, \bibinfo{pages}{762 -- 770},
  \doiprefix\url{10.1016/j.physa.2018.10.007} (\bibinfo{year}{2019}).

\bibitem{Kra:Red:03}
\bibinfo{author}{Krapivsky, P.~L.} \& \bibinfo{author}{Redner, S.}
\newblock \bibinfo{journal}{\bibinfo{title}{Dynamics of majority rule in
  two-state interacting spin systems}}.
\newblock {\emph{\JournalTitle{Phys. Rev. Lett.}}}
  \textbf{\bibinfo{volume}{90}}, \bibinfo{pages}{238701},
  \doiprefix\url{10.1103/PhysRevLett.90.238701} (\bibinfo{year}{2003}).

\end{thebibliography}

\section*{Acknowledgments}
This work has been partially supported by the National Science Center (NCN, Poland) through grants no. 2016/21/B/HS6/01256 and 2019/35/B/HS6/02530. 

\section*{Author contributions statement}
B.N. conducted extensive Monte Carlo simulations and analytical calculations for both versions of the model and wrote the original draft, B.S. conducted preliminary studies of the 3-state annealed version of the model consisting of the Monte Carlo simulations and analytical calculations for selected types of initial conditions, K.Sz-W. developed the model, designed and supervised the research. All authors reviewed and edited the manuscript.

\section*{Competing interests}
The authors declare no competing interests.

\end{document}